\documentclass[12pt]{article} 
\usepackage{epsfig} 
\usepackage{citesort} 
\usepackage{times} 

\newcommand{\eq}{\begin{equation}} 
\newcommand{\eqx}{\end{equation}} 
\newcommand{\eqn}{\begin{eqnarray}} 
\newcommand{\eqnx}{\end{eqnarray}}

\newcommand{\om}{\omega}

\newcommand{\nin}{\noindent}

\begin{document} 
\nin 
{\bf PACS Classification:} $79.20.Hx$, $82.53.Ps$, $61.80.-x$\\ \\ 
\begin{center} 
{\Large \bf ELECTRON CASCADES PRODUCED BY PHOTOELECTRONS IN DIAMOND }\\ 
\vspace{10mm} 

{\Large Beata~Ziaja $^{\ast,\,\dag,\,\ddag}$}, 
{\Large Abraham~Sz\"{o}ke $^{\ast,\,\S}$}, 
{\Large Janos~Hajdu $^{\ast}$} \footnote{e-mail:ziaja@tsl.uu.se, 
~szoke1@llnl.gov,~hajdu@xray.bmc.uu.se}\\ 

{\footnotesize 
\vspace{3mm} 
$^{\ast}$ \it Department of Biochemistry, Biomedical Centre, 
\it Box 576, Uppsala University, S-75123 Uppsala, Sweden\\ 
\vspace{3mm} 

$^{\dag}$ \it Department of Theoretical Physics, 
	 Institute of Nuclear Physics, 
\it Radzikowskiego 152, 31-342 Cracow, Poland\\ 
\vspace{3mm} 
$^{\ddag}$ \it High Energy Physics, Uppsala University, 
	 P.O. Box 535, S-75121 Uppsala, Sweden 
\vspace{3mm} 

$^{\S}$ \it Lawrence Livermore National Laboratory, Livermore, 
	 CA 94551, USA\\ 
} 
\end{center} 

\vspace{5mm} 
\nin 
{\bf Corresponding author:}\\ 
Janos Hajdu, Department of Biochemistry, Biomedical Centre,\\ 
Box 576, Uppsala University, S-75123 Uppsala, Sweden\\ 
Tel:+4618 4714449, Fax:+4618 511755, E-mail:hajdu@xray.bmc.uu.se \\ \\ 
\nin 

{\bf Abstract:} 
{\footnotesize 
Secondary electron cascades are responsible for significant ionizations in 
macroscopic samples during irradiation with X-rays. A quantitative 
analysis of these cascades is needed, e.g. for assessing damage in optical 
components at X-ray free-electron lasers, and for understanding damage in 
samples exposed to the 
beam. Here we present results from Monte Carlo simulations, showing the 
space-time evolution of secondary electron cascades in diamond. These cascades
follow the impact of a 
single primary electron at energies between $0.5-12$ keV, representing the 
usual range for photoelectrons. The calculations describe the secondary 
ionizations caused by these electrons, the three-dimensional evolution of the 
electron cloud, and monitor the equivalent instantaneous temperature of the free-electron 
gas as the system cools during expansion. The dissipation of the impact energy 
proceeds predominantly through the production of secondary electrons whose energies 
are comparable to the binding energies of the valence ($40-50$ eV) and the 
core electrons ($300$ eV) in accordance with experiments and the models of 
interactions. The electron cloud generated by a $12$ keV electron is 
strongly anisotropic in the early phases of the cascade ($t\le 1$ fs). At later 
times, the sample is dominated by low energy electrons, and these are scattered more 
isotropically by atoms in the sample. The results show that the emission of secondary 
electrons approaches saturation within about $100$ fs, following the primary impact. 
At an impact energy of $12$ keV, the total number of electrons liberated in the 
sample is $\le 400$ at $1000$ fs. The results provide an understanding of 
ionizations by photoelectrons, and extend earlier models on low-energy electron cascades 
($E=0.25$ keV, \cite{ziaja,ziaja2}) to the higher energy regime of the 
photoelectrons. 
} 
\vspace{6mm} 

In atomic or molecular samples exposed to X-ray radiation damage occurs. 
In light elements it proceeds mainly via the photoelectric effect. 
Photoelectrons and Auger electrons \cite{l1} are then emitted. 
They propagate through the sample, and cause further damage 
by excitations of secondary 
electrons. Photoelectrons released by X-rays of $\sim1$ \AA$\,$ wavelength 
($\sim12$ keV) are fast, 
$v\sim 660$ \AA/fs, and they 
can escape from small samples early in an exposure. 
In contrast, Auger electrons are slow ($v\sim95$ \AA/fs in carbon), so 
they remain longer in a sample, and it is likely that they will thermalize 
there. 
An analysis of electron cascades initiated by Auger electrons in diamond 
is described in \cite{ziaja2}. 
 
A detailed description of electron cascades initiated by an electron 
impact of energy between $\sim 0.5-12$keV is needed for a better 
understanding of radiation damage in larger samples as secondary 
ionization caused by 
propagating photoelectrons is significant there. Such processes need to be
investigated for planned experiments with free-electron-lasers (FEL).
Atomic and molecular clusters irradiated with VUV radiation have already 
been studied experimentally \cite{desy}.
 
Energetic photoelectrons ($E>0.5$ keV in carbon) propagate almost freely 
through the medium. They interact with single atoms, and this interaction 
is well described by the Born approximation \cite{l6prim,l6bis}. Here, we 
use the formalism of the Lindhard dielectric function \cite{l17prim} 
together 
with two optical models (the TPP-2 and Ashley's models 
\cite{l14,l14prim,l30,l15,l21,l21prim}) to describe the 
inelastic interactions of electrons with atoms. This approach takes into 
account the valence and core ionizations of an atom, following the 
inelastic scattering of a free electron. This approximation works 
well for energies up to about $10$ keV, and in this energy 
regime core ionization are responsible for not more than about $10$ \% of 
the total number of ionizations in solids \cite{pierop}. 
 
In this paper elastic scattering is treated in the muffin-tin potential 
approximation \cite{l6prim,l6bis,ziaja,ziaja2}. In previous studies on 
low-energy
electrons (with energies up to, 
$E=0.4$ keV), we used programs from the Barbieri/Van Hove Phase Shift 
package 
\cite{l12}. For electrons of energies, $E=0.4$ keV, and higher, we 
obtained elastic cross sections from the NIST database \cite{nist}. 

\begin{figure}[t]
\begin{center} 
\epsfig{figure=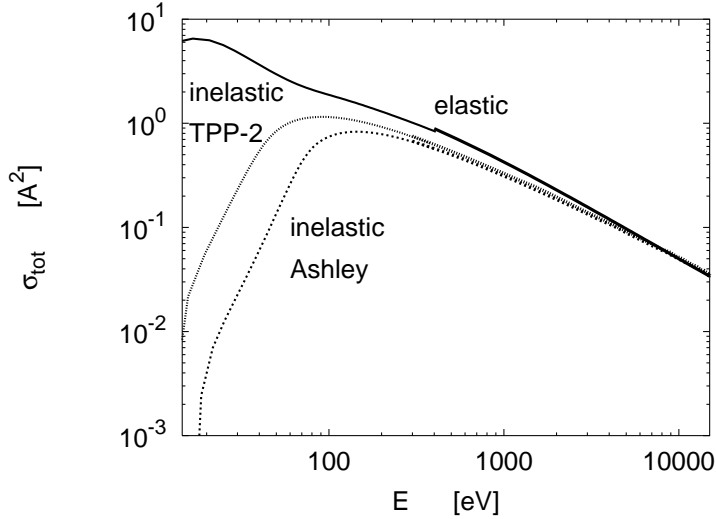,width=10cm}
\end{center} 
\caption{{\footnotesize Elastic and inelastic total cross sections for 
diamond.
Inelastic cross sections are obtained from the Lindhard approximation with 
core 
ionization taken into account using the TPP-2 optical model or 
Ashley's optical model. Elastic cross sections up to energies, $E=0.4$ 
keV, 
were derived with the Barbieri/Van Hove Phase Shift package \cite{l12}.
For larger energies, the elastic cross sections were taken from the NIST 
database.}
}
\label{f1} 
\end{figure}

Fig.\ \ref{f1} shows the total elastic and inelastic cross sections 
obtained from the calculations. The results show that for energies higher 
than $1$ keV, the elastic and inelastic cross 
sections are comparable, but for lower energies, $0.1<E<1$ keV, the 
elastic cross 
section is twice as large as the inelastic one. For very low energies, 
$E<0.1$ keV, 
the inelastic cross sections drop 
rapidly as the energy decreases, and elastic interactions become
predominant. This was true for both optical models (Ashley and TPP-2).

\begin{figure}[t] 
\begin{center}
a)\epsfig{width=10.5cm, file=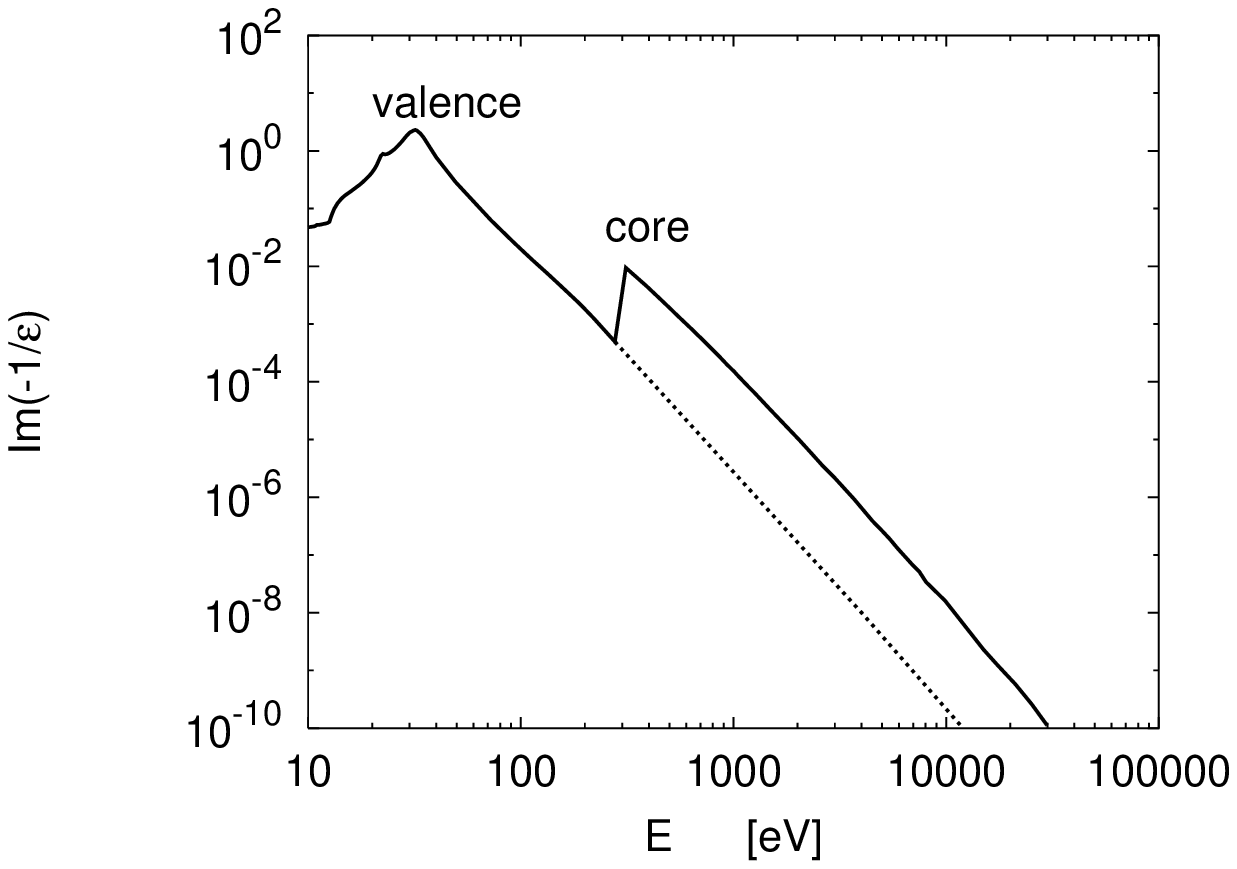}\\
b)\rule{10pt}{0pt}\epsfig{width=10cm, file=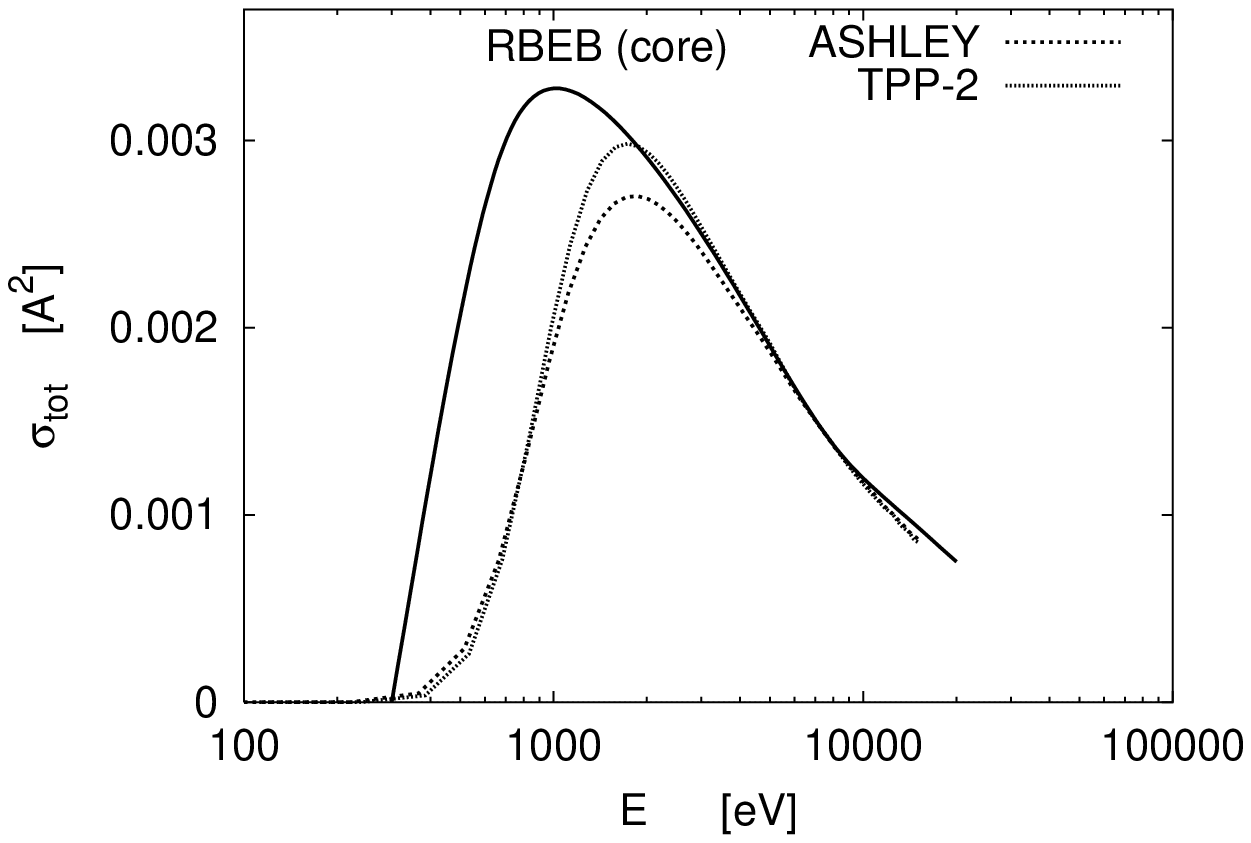}\\
\end{center}
\caption{{\footnotesize (a) Energy loss function of diamond, 
$Im(-1/\epsilon)$,
and (b) the total 
cross section for core ionization of diamond. Results from the Lindhard 
approximation, obtained with Ashley's and the TPP-2 models are compared to 
the prediction of  RBEB model for core ionization from K shell 
in carbon \cite{rbeb}. The binding energy of a K shell electron
in carbon is, $E_B\sim285$ eV.}
}
\label{f2} 
\end{figure}

Fig.\ \ref{f2} shows the energy loss function (ELF) and the cross 
sections for core ionization from the K shell of carbon as estimated 
from the Lindhard approximation, using the optical models above. The 
figures were compared to results from the relativistic binary-
encounter-Bethe model (RBEB) \cite{rbeb} for the total core ionization 
by impact electrons. This model was developed by combining a modified form 
of the Mott cross section and the leading dipole part of the Bethe cross 
section 
\cite{rbeb0}. Recently, it was extended to incident electrons with 
relativistic 
energies \cite{rbeb}. When the incoming electron is fast, we can use the 
Bethe-Fermi approximation \cite{l16,bethe}. It replaces the electric field 
of the incoming electron by an electromagnetic pulse of the same, short 
duration. Impact ionization is then proportional to the dipole transition 
probability caused by the short, non-periodic electromagnetic pulse. 
Incidentally, that is the basis of the two optical models used here. The 
ionization probability is proportional to the overlap of the bound 
electron wave function with that of the secondary electron. Simple 
scaling considerations predict that if valence electrons are ionized, 
their kinetic energies should not be much larger than a few times the 
shell binding energy. Figure 3 shows that the mean energy of the 
secondary electrons in diamond is about 40-50 eV (close to the energy 
of L shell electrons), reaching peak energies of about $300$ eV (close 
to the energy of core electrons). The energy loss function in Fig.\ 
\ref{f2} is 
the sum of the large valence and small core contributions which add to the 
valence 
contribution at energies, $E\sim 0.3$ keV. For comparison, the binding 
energy of 
the K shell in carbon is, $E_B\sim285$ eV. It is difficult to separate the 
contribution of the core excitations from the valence excitations on the 
base of the 
ELF alone. Therefore, we make here a rough estimate of the pure core 
contribution 
by subtracting the valence component from ELF. The valence component was 
extrapolated at the edge of the core peak, i.\ e.\ for energies larger 
than $277$ eV 
(cf.\ Fig.\ \ref{f2}). The cross section obtained from this core 
contribution 
underestimates the RBEB cross section at energies smaller than $1-2$ keV. 
This is 
due to the fact that the optical approximation does not apply at the core 
threshold. 
At larger energies our estimate agrees well with the RBEB predictions. 
This 
discrepancy does not make our results inaccurate. 
 
\begin{figure}[t] 
\begin{center}
\epsfig{width=10cm, file=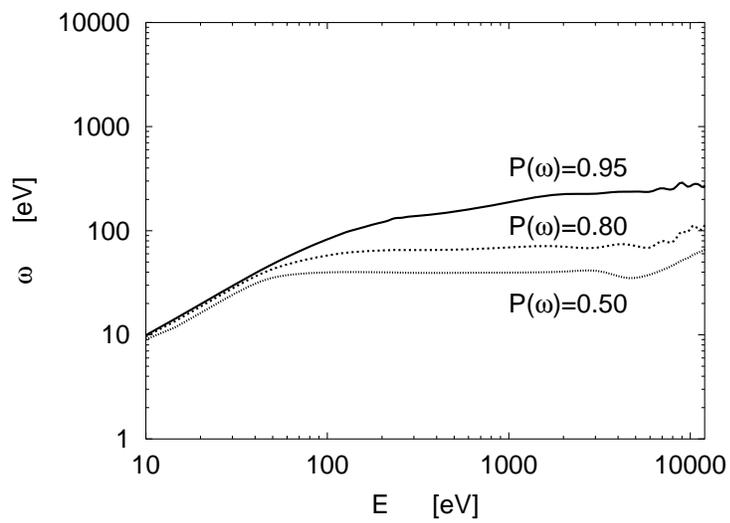}\\
\end{center}
\caption{{\footnotesize Energy loss, $\om$, in a single inelastic 
scattering event during electron-atom interactions in diamond as a 
function 
of electron energy. Plots show results at fixed integrated probabilities, 
$P(\om|E)=0.5,0.8,0.95$, obtained with the TPP-2 model. The probability 
$P(\om|E)$ is the integrated probability that the energy loss in a 
scattering 
is less or equal $\omega$. 
Results from Ashley's model were very similar (not shown)}
}
\label{f33} 
\end{figure}

Fig.\ \ref{f33} shows the energy loss, $\om$, during a single inelastic 
scattering 
of an electron on an atom as a function of the electron energy. 
These are the values obtained at fixed integrated probabilities, 
$\displaystyle P(\om|E)=\left[ \int_{\om_{min}}^{\om}\,d\om^{\prime}(d\sigma(E)/
d\om^{\prime})\right] /
\left[ \int_{\om_{min}}^{\om_{max}} \,d\om^{\prime} 
(d\sigma(E)/d\om^{\prime})\right]$, 
and represent the integrated probabilities of inelastic electron-atom 
scattering 
with the energy loss that is less or equal to $\om$ at the impact energy, 
$E$. 
The energy loss, $\om$, does not exceed $0.3$ keV, and its value becomes 
independent from the impact energy, $E$, at energies greater than $1$ keV. 
This is expected as one-electron excitations are 
predominant in inelastic scatterings \cite{l16,bethe}.

\subsection*{Results}

We performed a large set of Monte Carlo simulations in diamond, showing 
the path of an impact electron and the secondary electron cascade 
triggered by the 
electron. Electron trajectories 
were simulated as described in \cite{ziaja2} in such a way that no energy 
loss was 
permitted to the lattice. This approximation gave a better 
estimate of ionization rates in \cite{ziaja2} than the approximation where 
the energy loss to the lattice was allowed. The starting position 
of the impact electron at $t=0$ fs was at the origin {\bf x}=(0,0,0) of an 
arbitrarily chosen coordinate system, and the velocity of the impact 
electron was along the Z-axis. Motions of holes and ionizations
by the holes were neglected since they influence the dynamics of electrons
at very low electron energies only.
The space-time characteristics of secondary cascades were recorded as 
functions of the impact energy and time. 

The energy dependence of the elastic and inelastic cross sections had a 
strong influence on the dynamics of the electron cloud (cf.\ Fig.\ 
\ref{f1}). At longer times, when most of the electrons have already cooled 
to low energies, 
elastic (isotropic) scatterings dominate the sample. Electrons then 
propagate randomly through the sample, and their distibution is isotropic. 
Earlier studies show similar behaviour for Auger electrons \cite{ziaja2}.

\begin{figure}[t]
a)\epsfig{width=8cm, 
file=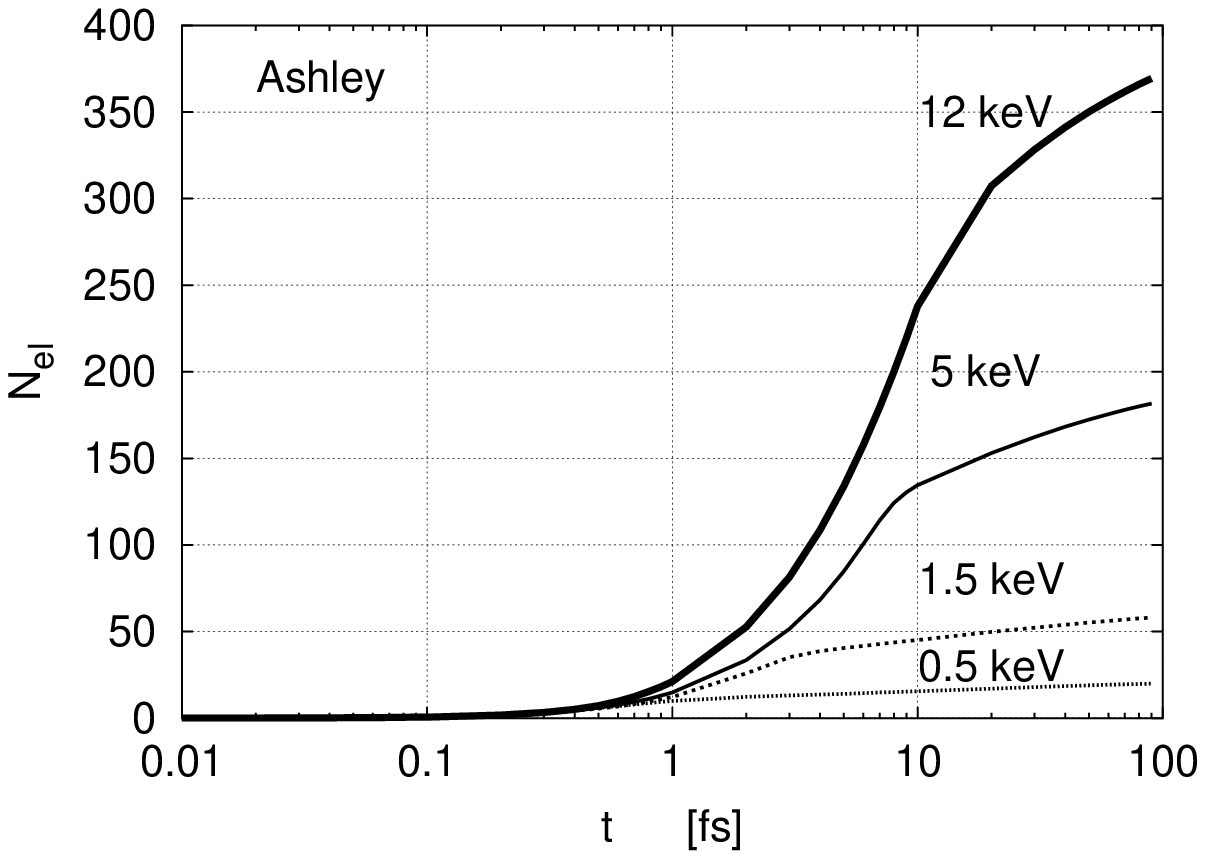}
\epsfig{width=8cm, 
file=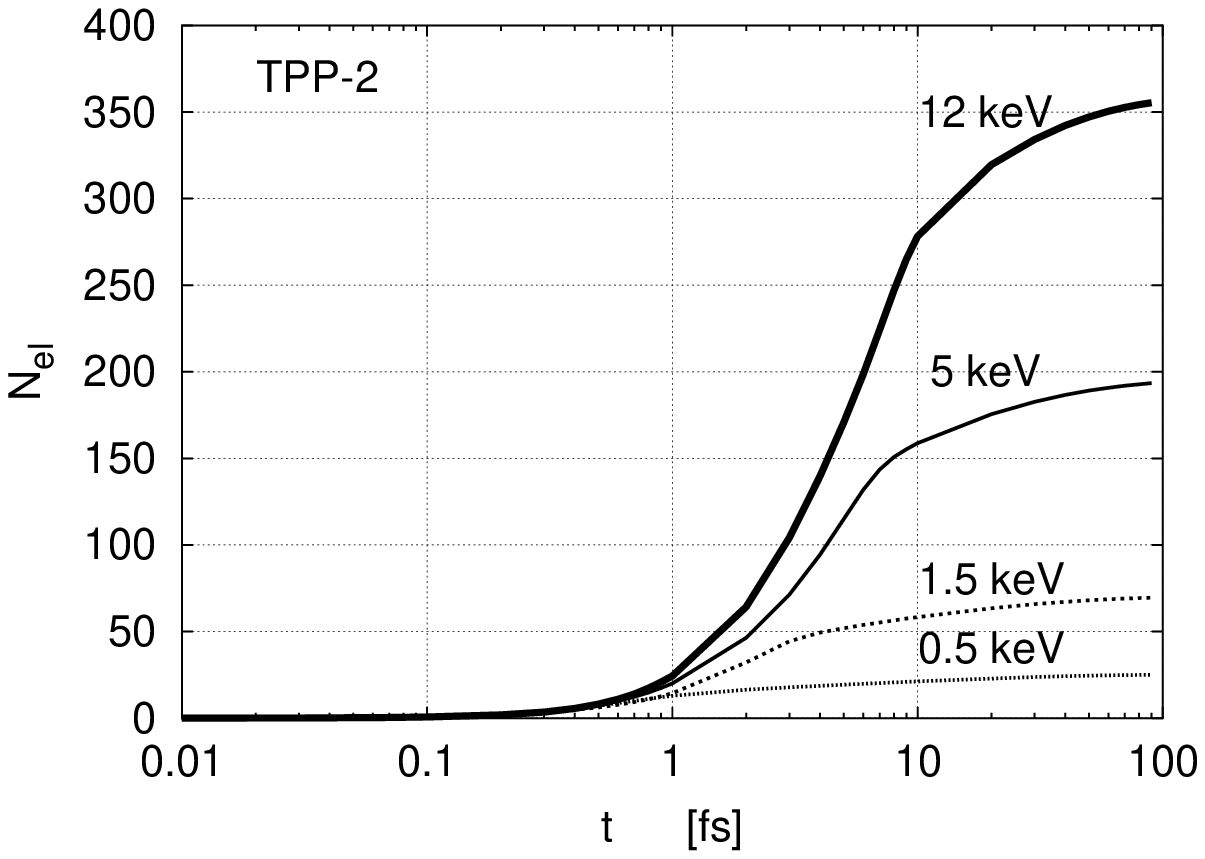}\hfill\mbox{}\\
b)\epsfig{width=8cm, 
file=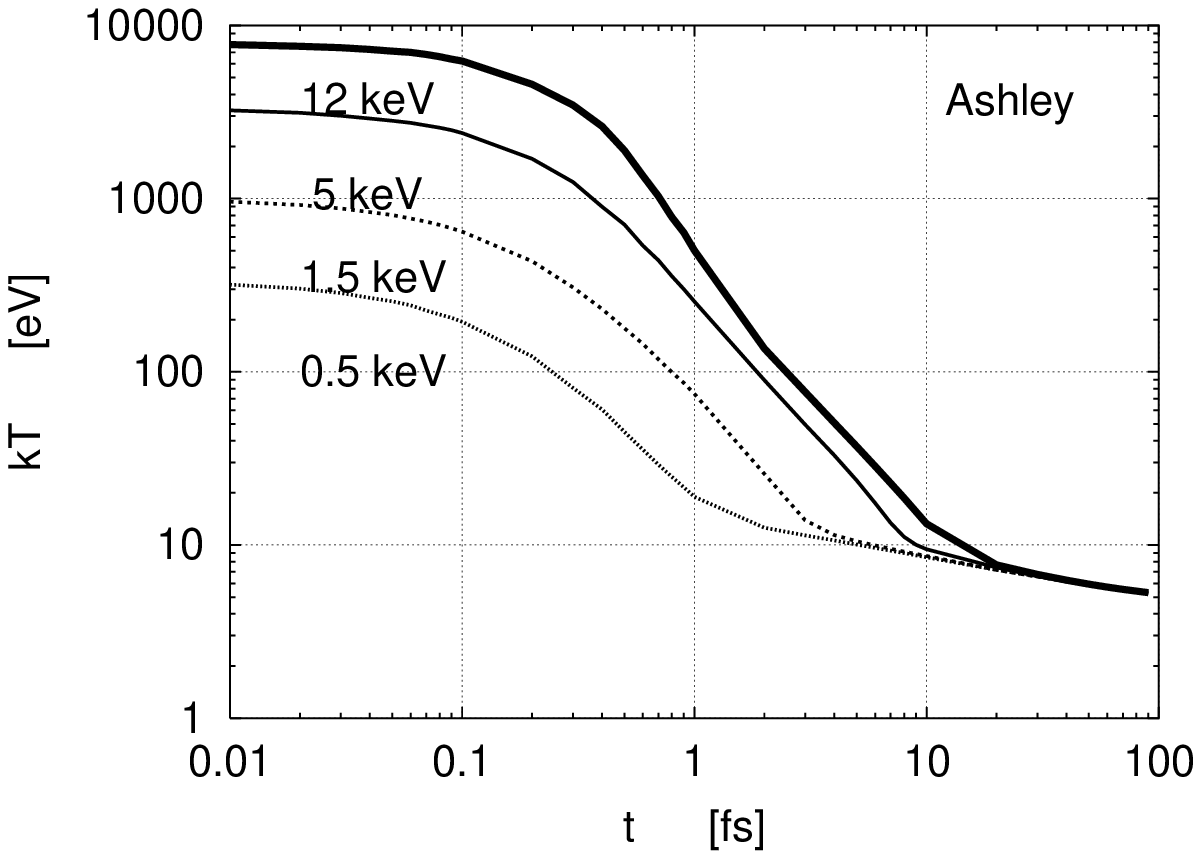}
\epsfig{width=8cm, 
file=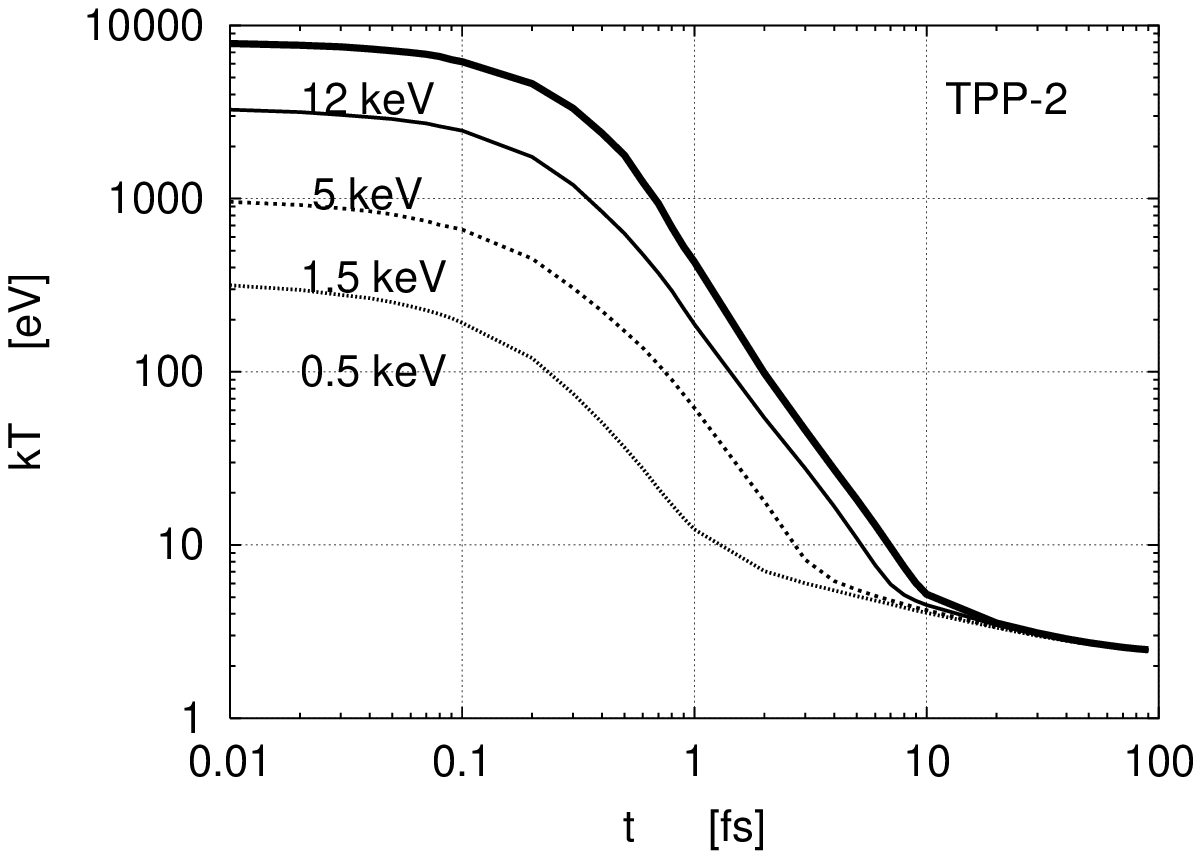}\hfill\mbox{}\\
\caption{{\footnotesize {\bf (a)} Average number of secondary electrons emitted, $N_{el}$, vs. time;
{\bf (b)} The equivalent instantaneous temperature $kT$ of electron gas vs. time 
averaged over $500$ cascades. Curves correspond to the results obtained 
at different electron impact energies $E=0.5-12$ keV from Ashley's model 
and the TPP-2 model with {\bf no} energy transfer allowed to the lattice.}
}
\label{f3}
\end{figure}

{\bf Evolution} of the cascade was analysed through {\bf (a)} 
the number of secondary ionizations, $N_{el}(t)$, and {\bf (b)} 
the equivalent instantaneous temperature of the free electron gas $kT(t)$.
Quantities (a), (b) were averaged over a number of cascades. Figure 
\ref{f3} shows the results obtained with Ashley's and the TPP-2 
optical models for impact energies of $E=0.5,1.5,5,12$ keV. $0.5$ keV 
corresponds to the lowest energy of an incident electron at which the Born 
approximation may be applied when calculating the electron-atom cross 
sections in carbon \cite{models3}.

The number of ionization events within the first femtosecond is
within 10-25 for electron impact energies of $E=0.5-12$ keV. For impacts 
at lower energies, $E\le 1.5$ keV, the number of ionizations was between 
$20-70$, and saturated within $100$ fs. At higher impact energies there 
was an increase in the number of the ionization events after $t>100$ fs, 
however, the 
sample was then dominated by low energy electrons ($E<20$ eV) 
(cf.\ Figs.\ \ref{f5}), and thus not many ionizations could be expected
to occur. We checked that at $1000$ fs, there was a total of about $400$ 
ionizations induced by an electron impact of $12$ keV (Ashley's model), 
compared to $370$ ionizations at $100$ fs. When using the TTP-2 model, the 
corresponding 
numbers were $\sim360$ both at $100$ fs and $1000$ fs at $12$ keV impact 
energy.

The equivalent instantaneous temperature of the electron gas decreased
as the cascade evolved. We used this temperature equivalent since the 
electron gas was far from thermal equilibrium. However, the equivalent
temperature is still a quantity conserved in electron-electron collisions.
It is worth noticing that after $\sim 10$ fs the temperature curves 
obtained at totally different primary energies were very similar. 
This indicated that the number of electrons
was not much influenced by the energy of the primary electron but rather
by secondary electrons of lower energies ($<60$ eV) which dominated the 
sample 
after $10$ fs. At $100$ fs the temperature of the electron gas dropped to 
$\sim 5$ eV (Ashley's) or $\sim 2.5$ eV (the TPP-2 model).

\begin{figure}[t]
a)\epsfig{width=8cm, 
file=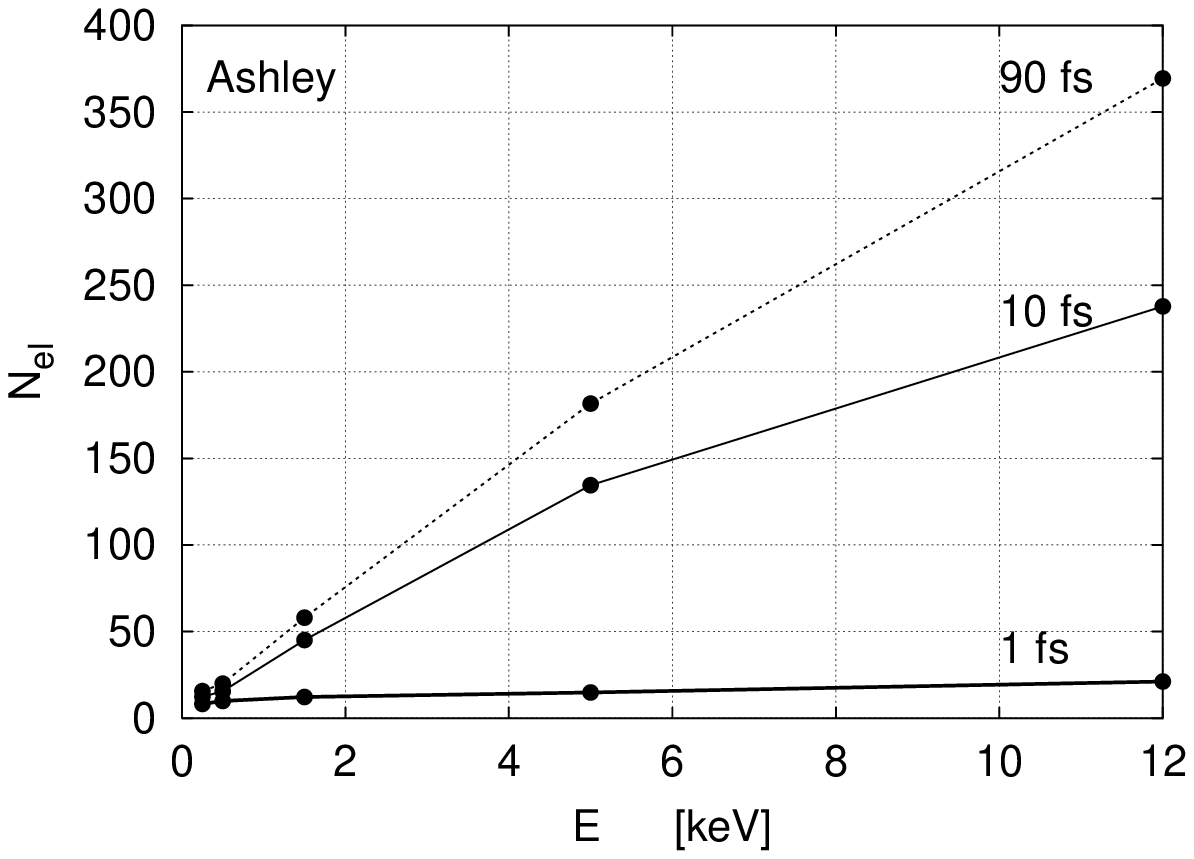}
\epsfig{width=8cm, 
file=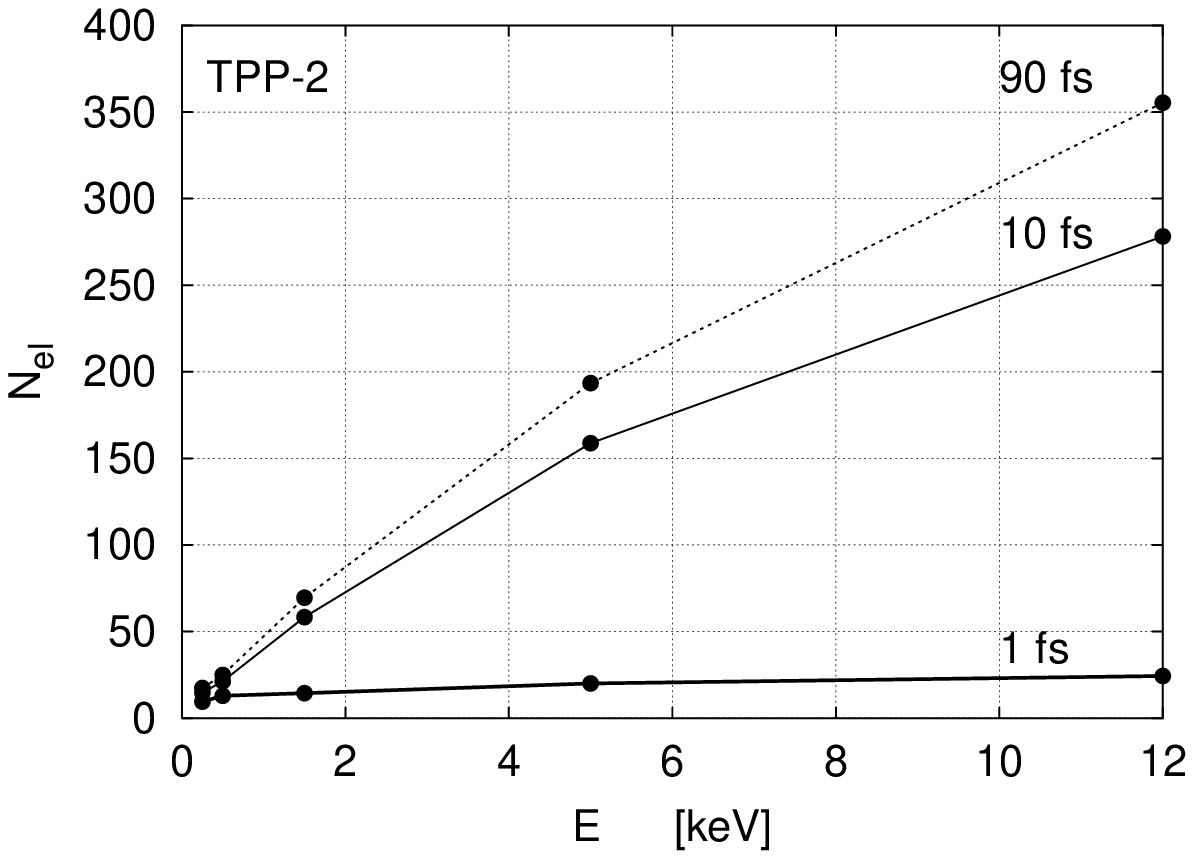}\hfill\mbox{}\\
b)\epsfig{width=8cm, 
file=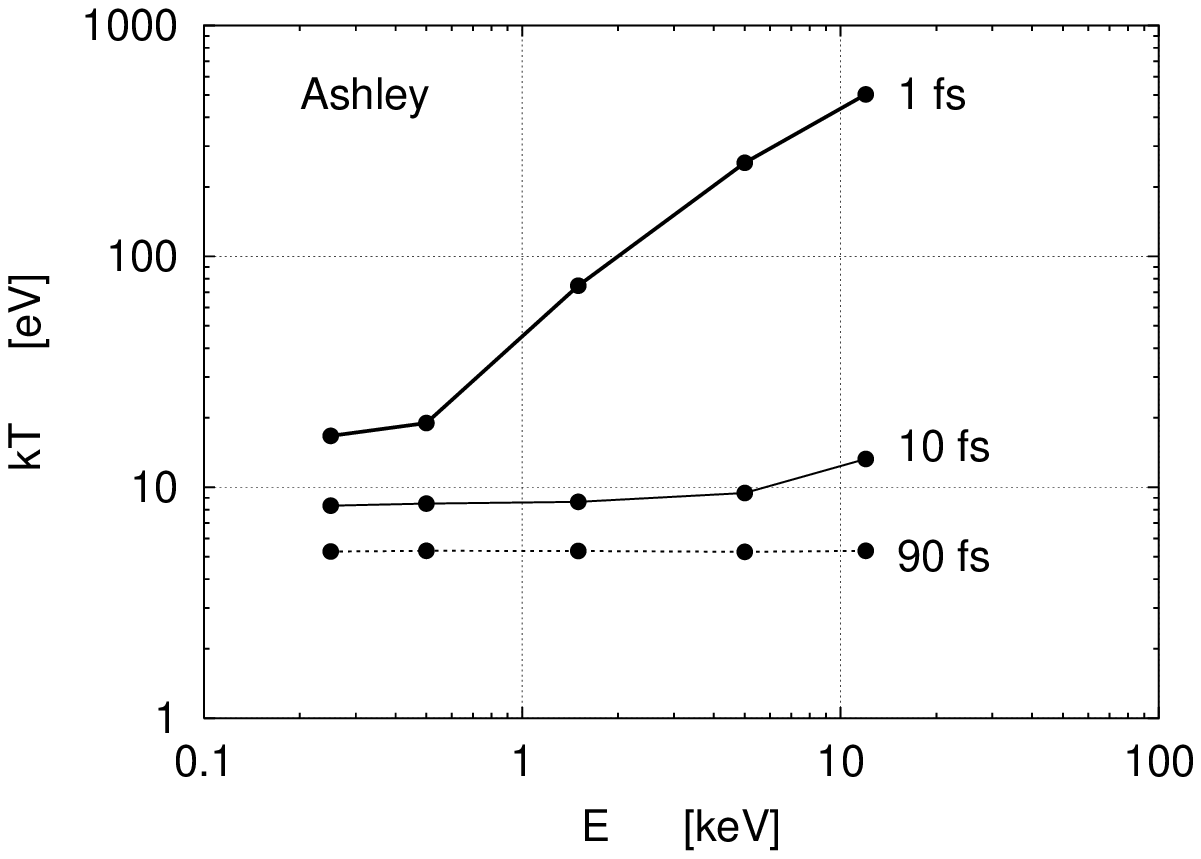}
\epsfig{width=8cm, 
file=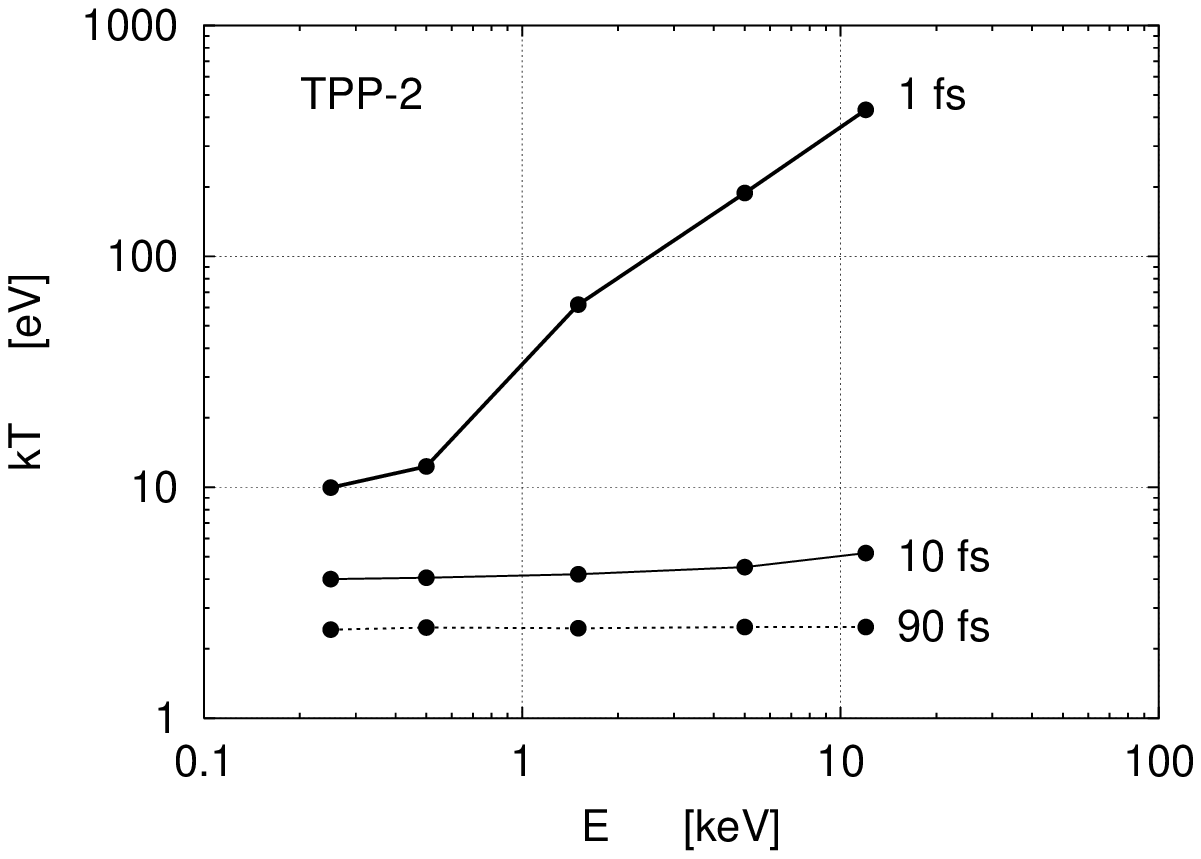}\hfill\mbox{}\\
\caption{{\footnotesize {\bf (a)} Average number of secondary electrons emitted, $N_{el}$, vs. energy, $E$;
{\bf (b)} The equivalent instantaneous temperature $kT$ of electron gas 
vs. energy, $E$ averaged over $500$ cascades. Curves correspond to the 
results obtained at different times $t=1,10,90$ fs from Ashley's model 
and the TPP-2 model with {\bf no} energy transfer allowed to the lattice.
The data were collected at primary energies of $E=0.25,0.5,1.5,5,12$ keV. 
The results at energy $E=0.25$ keV were taken from \cite{ziaja2}}
}
\label{f4}
\end{figure}

\begin{figure}[t]
a)\epsfig{width=8cm, 
file=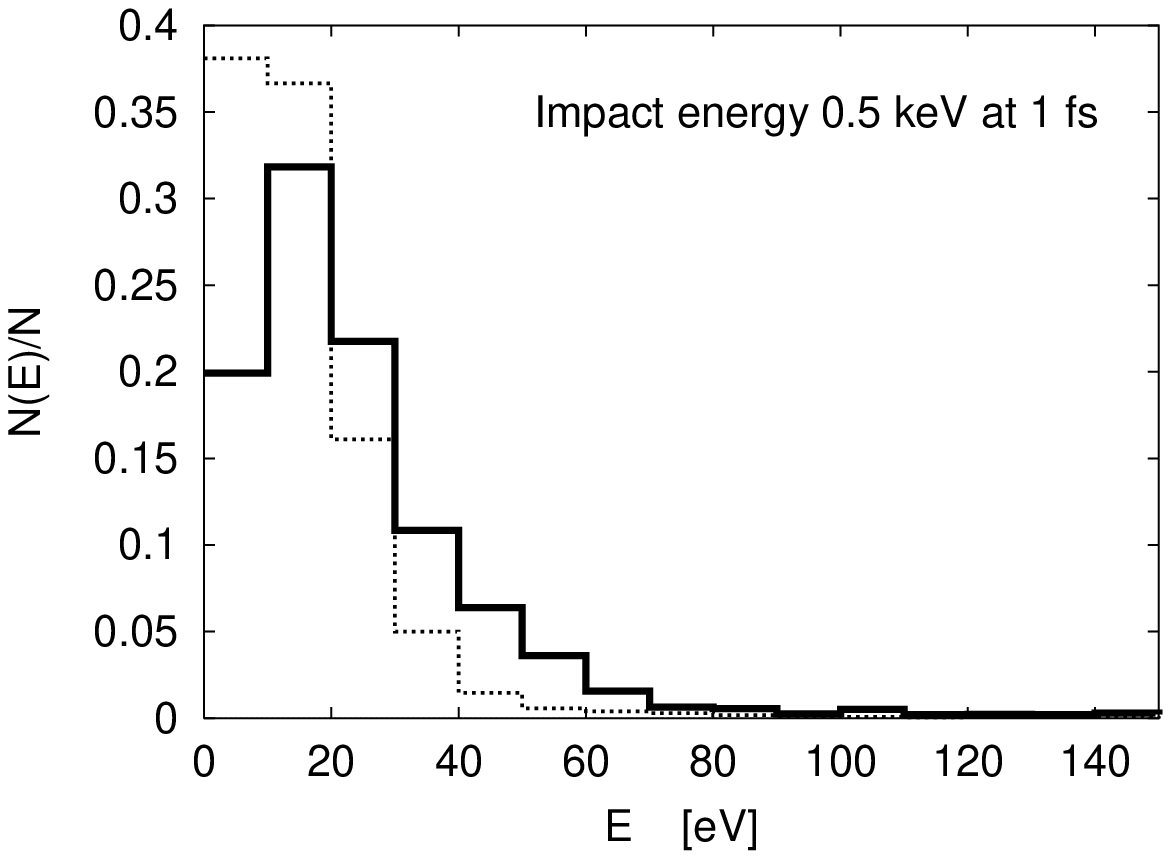}
\epsfig{width=8cm, 
file=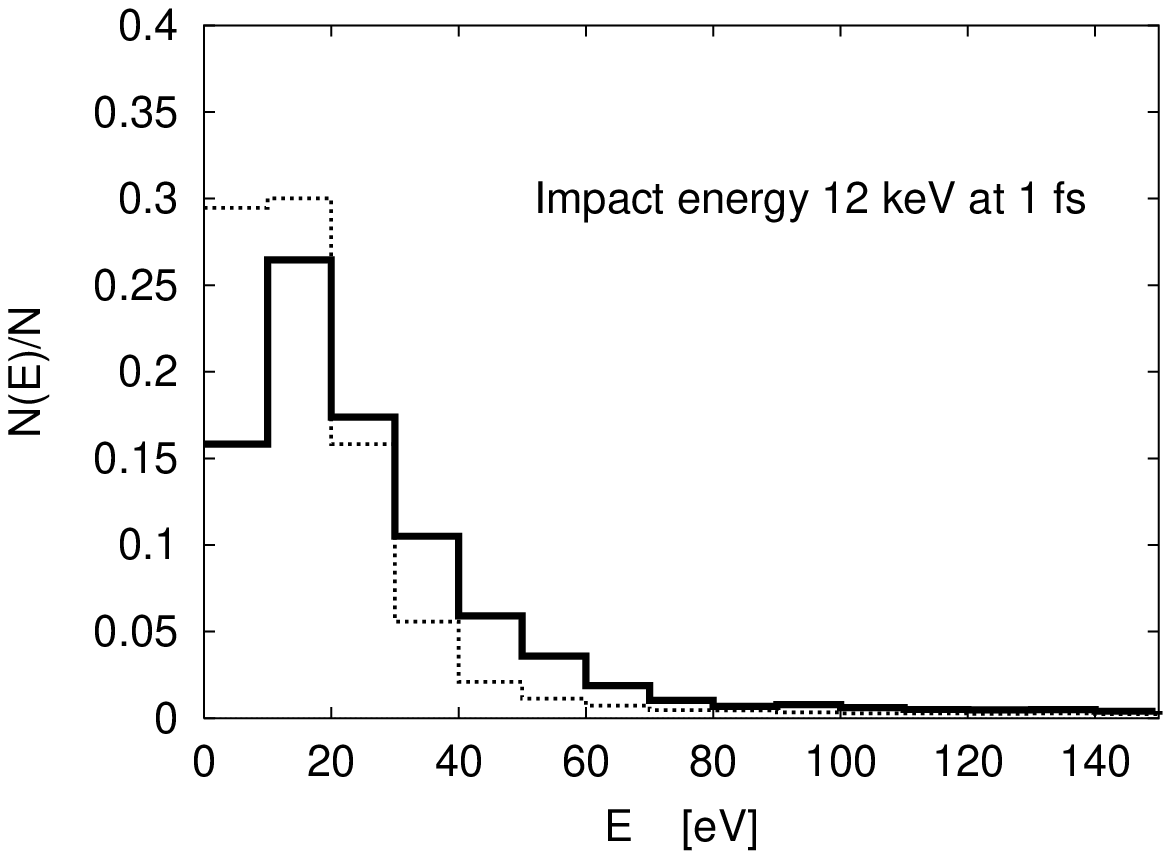}\hfill\mbox{}\\
b)\epsfig{width=8cm,file=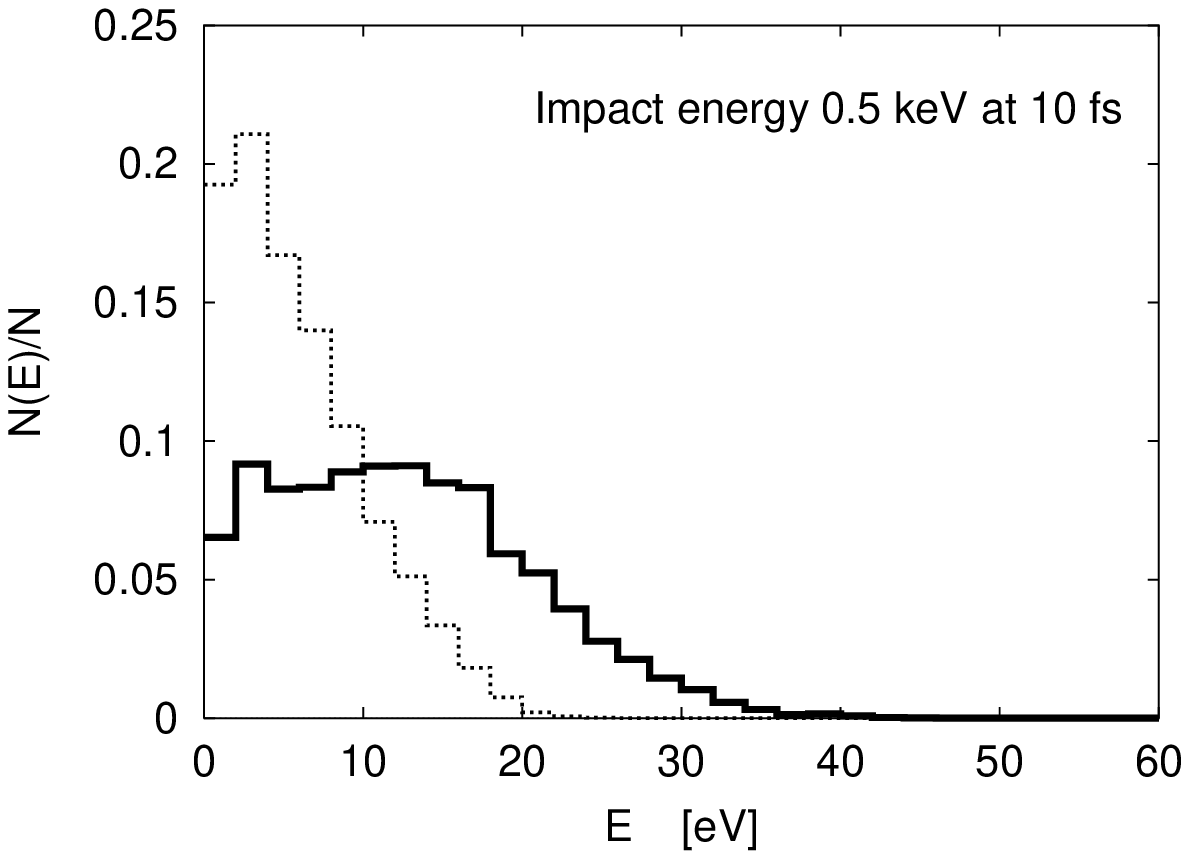}
\epsfig{width=8cm,file=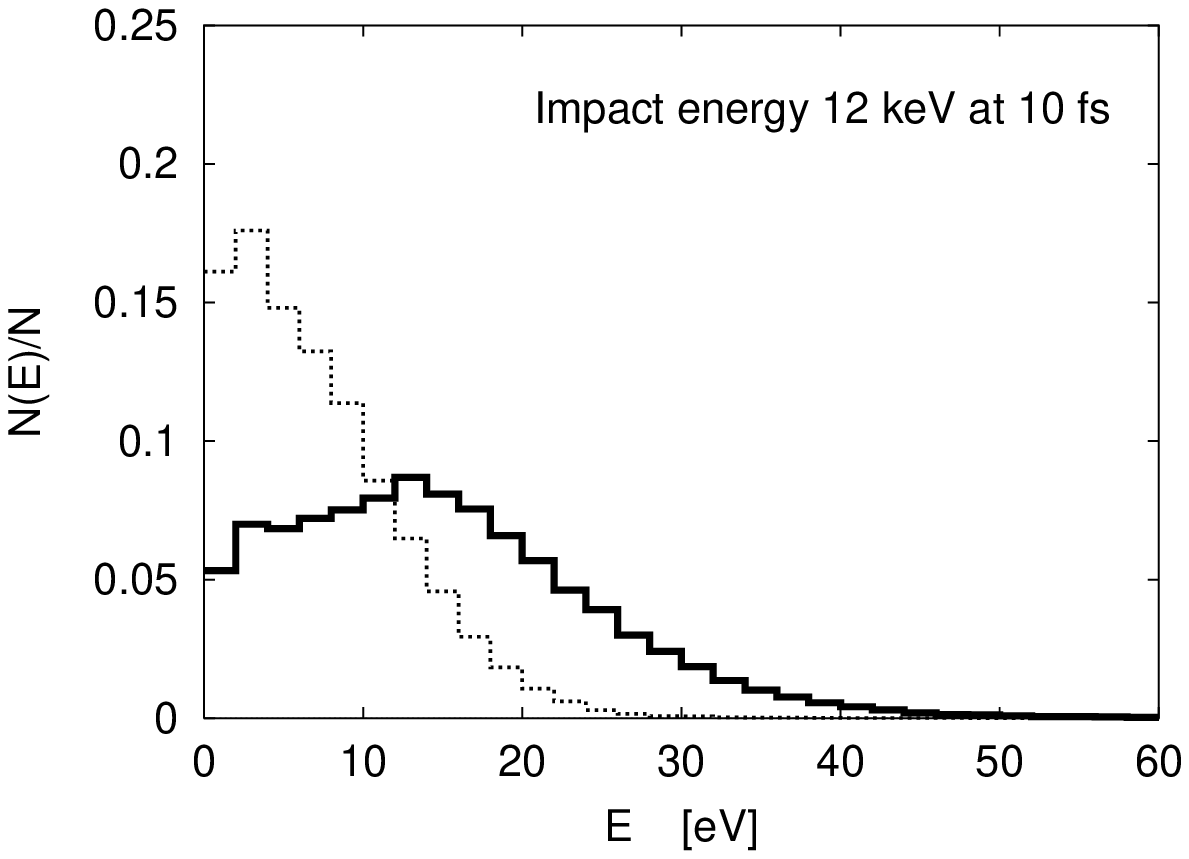}\hfill\mbox{}\\
c)\epsfig{width=8cm,file=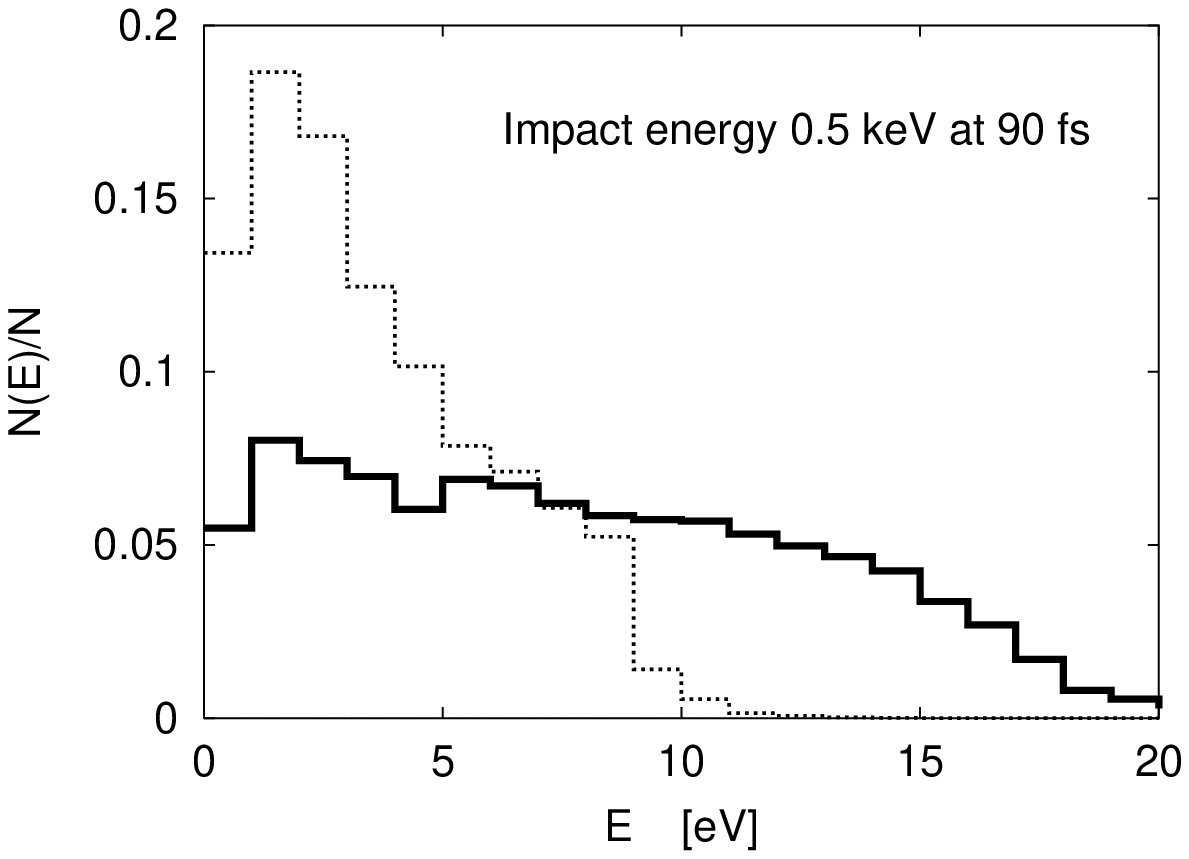}
\epsfig{width=8cm,file=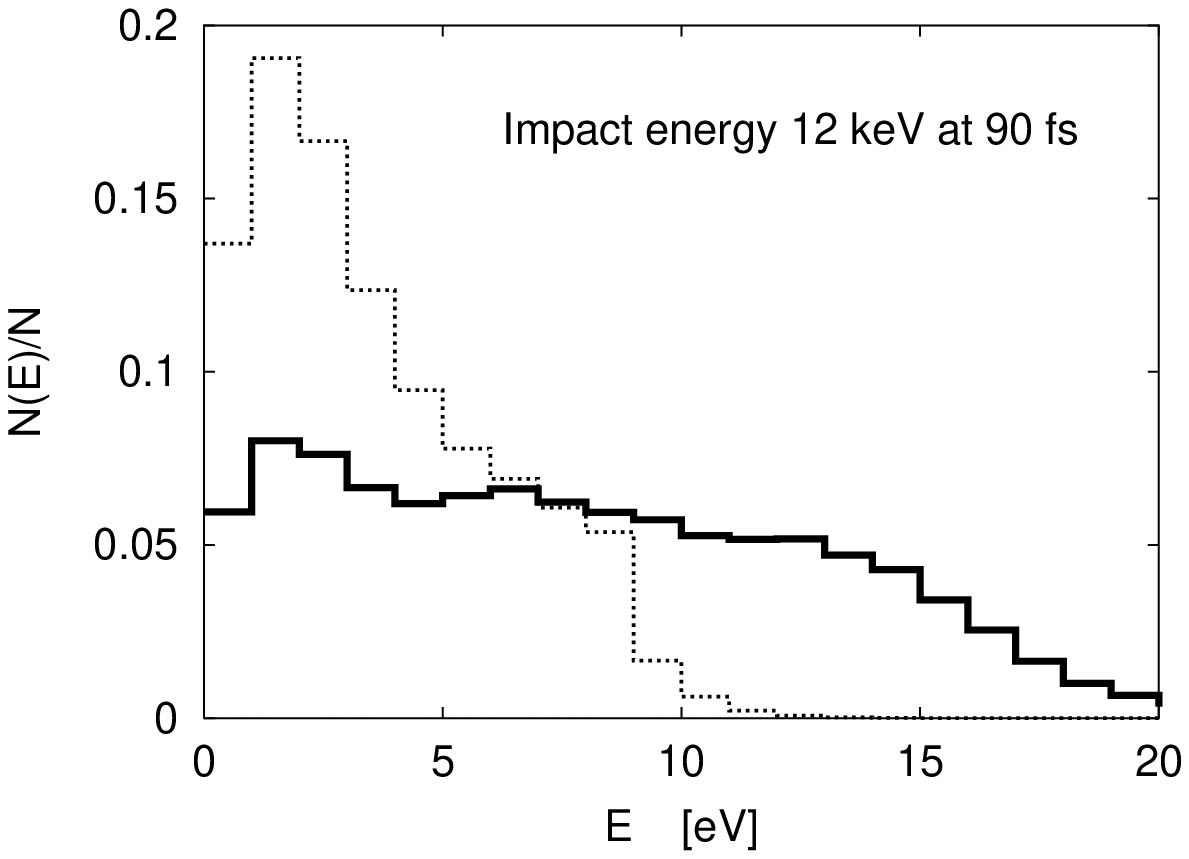}\hfill\mbox{}\\
\caption{{\footnotesize Energy distribution, $N(E)/N$, (fraction of 
electrons
per bin) among electrons (histogram) 
at {\bf (a)} $t=1$ fs; {\bf (b)} $t=10$ fs; and {\bf (c)} $t=90$ fs. 
Histograms correspond to results obtained at electron impact of $E=0.5$ 
keV 
(left) and $E=12$ keV (right) from Ashley's model (solid line) and the TPP-2 
model (dotted line) when {\bf no} energy transfer to the lattice is allowed.
}}
\label{f5}
\end{figure}

Fig.\ \ref{f4} shows plots of {\bf (a)} the average number of electrons 
released, $N_{el}$; {\bf (b)} the equivalent instantaneous temperature, 
$kT$, of the electron gas as a function of the energy, $E$, of the primary 
electron.
These curves describe results obtained at different times 
($t=1,10,90$ fs), and were based on Ashley's model and on 
the TPP-2 model with {\bf no} energy transfer allowed to the lattice. The 
data represent primary energies of $E=0.25,0.5,1.5,5,12$ keV. Results for 
impact energy $E=0.25$ keV were taken from Ref.\ \cite{ziaja2}. The 
results can be used for the interpolation of the number of ionizations, 
and the temperature of the 
electrons at energies ranging from $E=0.25$ keV to $E=12$ keV. 
 
Fig.\ \ref{f4} shows that the number of secondary electrons, $N_{el}$, is 
approximatly proportional to the impact energy. The slight curvature of 
the 
dependence indicates that the system has not yet reached equilibrium at 
$90$ fs after the primary impact (cf. Fig.\ \ref{f3}). 
 
{\bf The energy distribution of secondary electrons}. The positions and 
velocities 
of electrons recorded at times, $t=1,\,10,\,90$ fs at 
energies of $E=0.5,1.5,5,12$ keV were collected from all cascades, 
and put into one file. Using these data, histograms for the energy 
distributions 
were obtained, $N(E)/N$, at these time points. Number,
$N(E)=\sum_{i=1}^{500}\,N_i(E)$, is the average number of electrons in a 
bin,
$(E,E+\Delta E)$, averaged over a number of cascades. Correspondingly, 
$N_i(E)$ was the number of electrons found in that bin for the ith 
cascade. 
These distributions were normalized to the total number of electrons, 
$N=\sum_{E}\,N(E)$. Fig.\ \ref{f5} shows the histograms at impact energies
of $E=0.5$ keV and $E=12$ keV. As expected, the energy histograms show 
that 
the number of low-energy electrons
increased with time. One may notice that the dissipation of the impact 
energy was fast. At $1$ fs, most electrons had energies lower
than $300$ eV in the cascade. This follows from the assumption that 
one-electron
excitations dominate electron-atom interactions, and events in which 
energy transfer exceeds the threshold for core 
ionizations by a secondary electron are rare.
At $10$ fs most of the electrons had energy lower than $40$ eV in all 
cascades. 
The overall low energy of the electrons at that time influence 
ionization rates and the expansion rate of the electron cloud, and these 
both slowed 
down after $10$ fs. At $100$ fs there were only low energy electrons 
($E<20$ eV) present 
in the sample, and elastic scatterings dominated the dynamics of the 
electron 
cloud.

\begin{figure}[t]
\epsfig{width=8cm, 
file=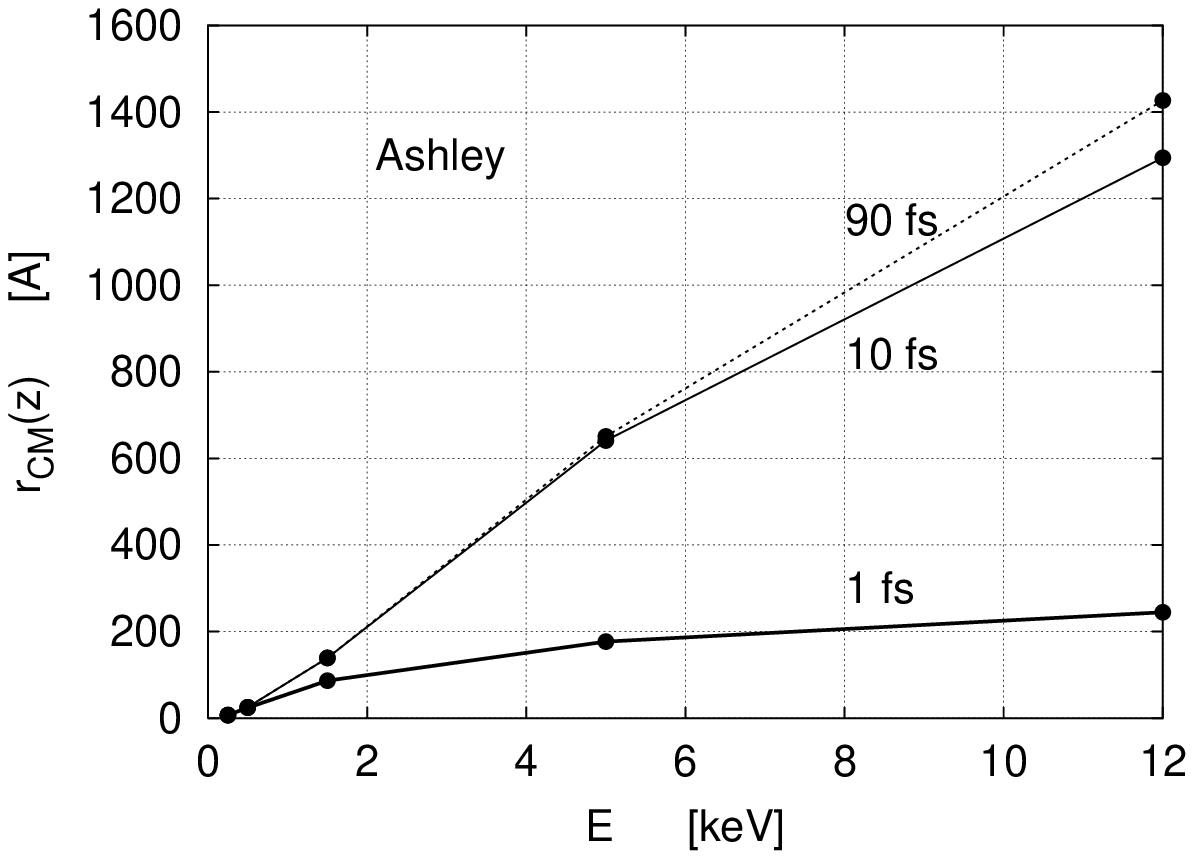}
\epsfig{width=8cm, 
file=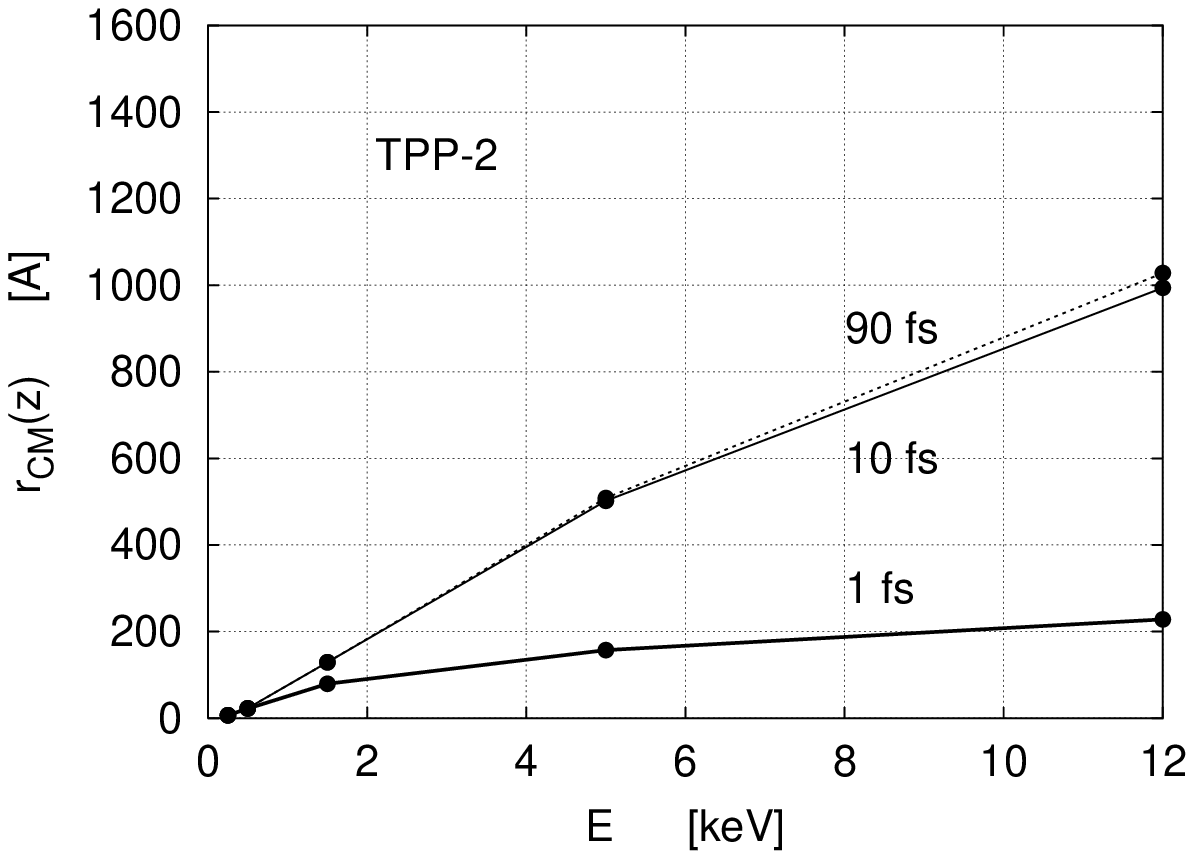}\hfill\mbox{}\\
\caption{{\footnotesize Shifts of the center of mass of the cloud in the 
direction of the primary impact (Z-axis) vs. energy, $E$, in respect to 
the point of emission of the primary impact. Data on the electron cloud 
were collected from 500 cascades. Curves correspond to the 
results obtained at different times $t=1,10,90$ fs from Ashley's model 
and the TPP-2 model with {\bf no} energy transfer allowed to the lattice 
at primary energies of $E=0.5,1.5,5,12$ keV. }
}
\label{f10}
\end{figure}

{\bf Spatial distribution of secondary electrons}. In order to describe 
the spatial distribution of the secondary electron cloud in cascades 
triggered by $E=0.5$ and $E=12$ keV electrons, results from 500 simulations 
were analysed at these energies. 
Most electrons lay within a sphere whose radius depends on
the electron impact energy. At 1 fs, it was $r_{cloud}\sim 50$ \AA$\,$ for 
$0.5$ keV impact energy, and $r_{cloud}\sim 600$ \AA$\,$ for $12$ keV impact 
energy. At $100$ fs the corresponding radii were between $150$ \AA$\,$ and 
$3100$ \AA$\,$ respectively. At this time, the center of mass 
of the cloud moved $23-1400$ \AA$\,$ away from the starting point in the 
direction of the primary impact at energies of $0.5-12$ keV, respectively
(Fig.\ \ref{f10}). The shifts in other directions were small 
($1-33$ \AA).

The sphericity tensor, 
$\displaystyle S^{ab}=\left[\sum_{i=1}^{N}\,r_i^a r_i^b\right]/
\left[\sum_{i=1}^{N}\,r_i^2\right]$, calculated for the collective cloud 
was almost diagonal. 
Non-diagonal elements were $\sim 10^1-10^4$ times smaller than the 
diagonal ones. 
The sphericity tensor was diagonalized, and the sphericity parameter, $S$, 
was obtained from its eigenvalues: $S=3/2\,(\lambda_2+\lambda_3)$, 
where $\lambda_1 \geq \lambda_2 \geq \lambda_3$. For electron impacts
of $0.5$ and $12$ keV $S$ falls within the interval of $0.91-0.96$ at $10$ 
fs and $0.91-0.98$ at $90$ fs, suggesting that the spatial distribution of 
the collective cloud was practically isotropic at that stage. 

Strong differences at the spatial distribution  of the electron cloud 
manifested at $1$ fs. At low impact energies ($0.5$ and $1.5$ keV), 
$S$ was within the interval of $0.8-0.9$, while at $5$ keV $S$ was 
$0.4-0.5$, and at $12$ keV, $S$ was around $0.30-0.35$. This shows a 
strong anisotropy with increasing impact energy. The distribution of the 
cloud was cylindrically symmetrical along the primary impact vector but it 
was elongated along the direction of the impact. 
This is due to the fact that at $1$ fs, the primary electron was 
much faster than the secondary electrons. At later times the 
sample became dominated by low energy electrons, which were scattered more 
isotropically. This is reflected in the progressing isotropy in the 
distribution. 
 
\begin{figure}[t]
\begin{center}
\epsfig{width=8cm, file=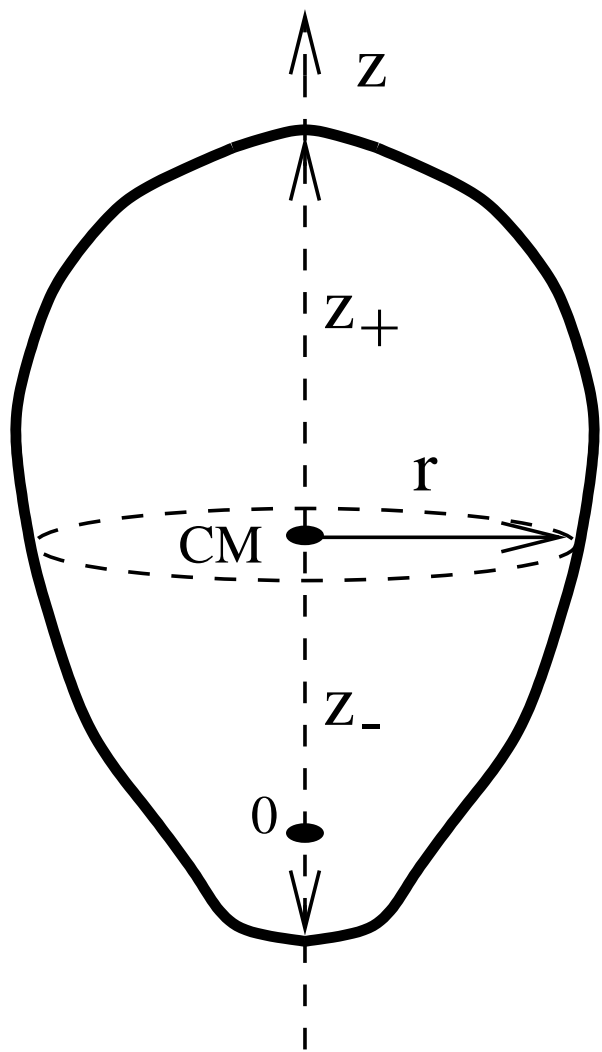}
\caption{{\footnotesize Schematic plot of the electron cloud.}}
\label{f11}
\end{center}
\end{figure}

\begin{figure}[t]
a)\epsfig{width=8cm, 
file=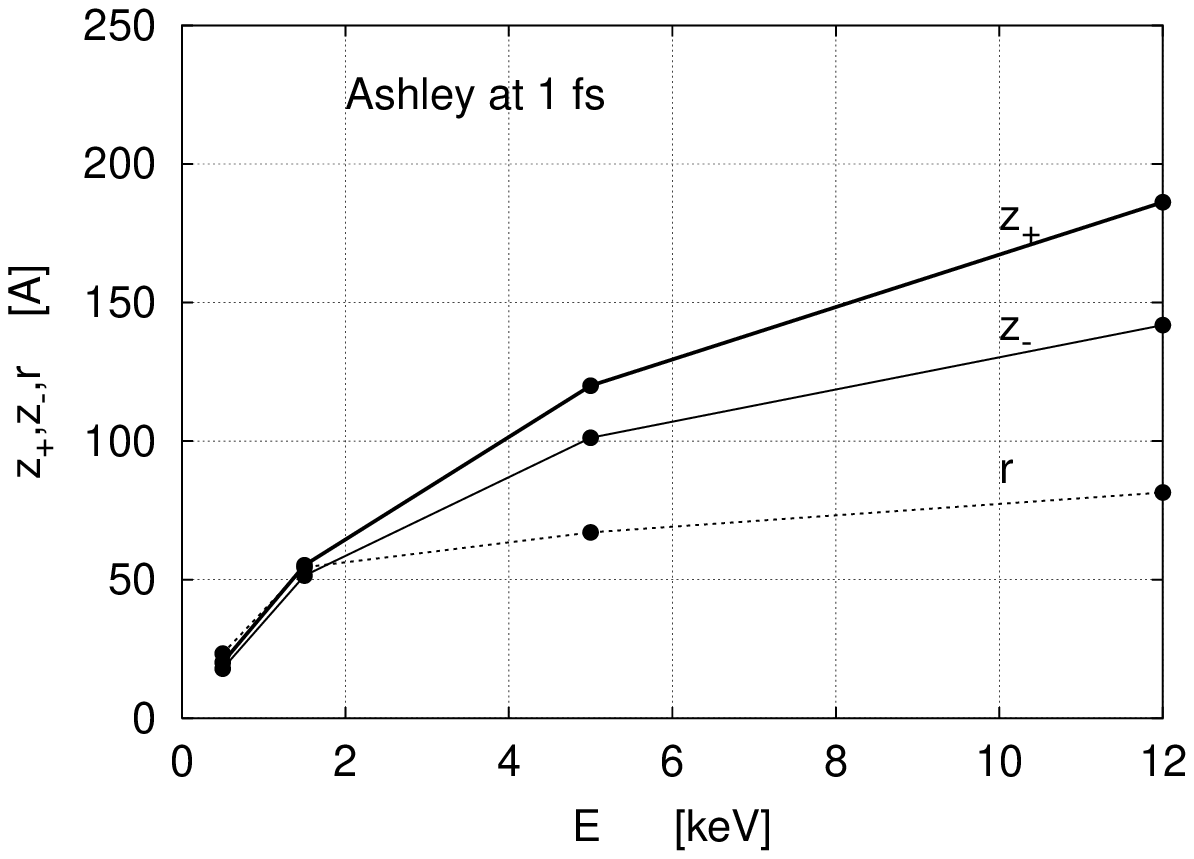}
\epsfig{width=8cm, 
file=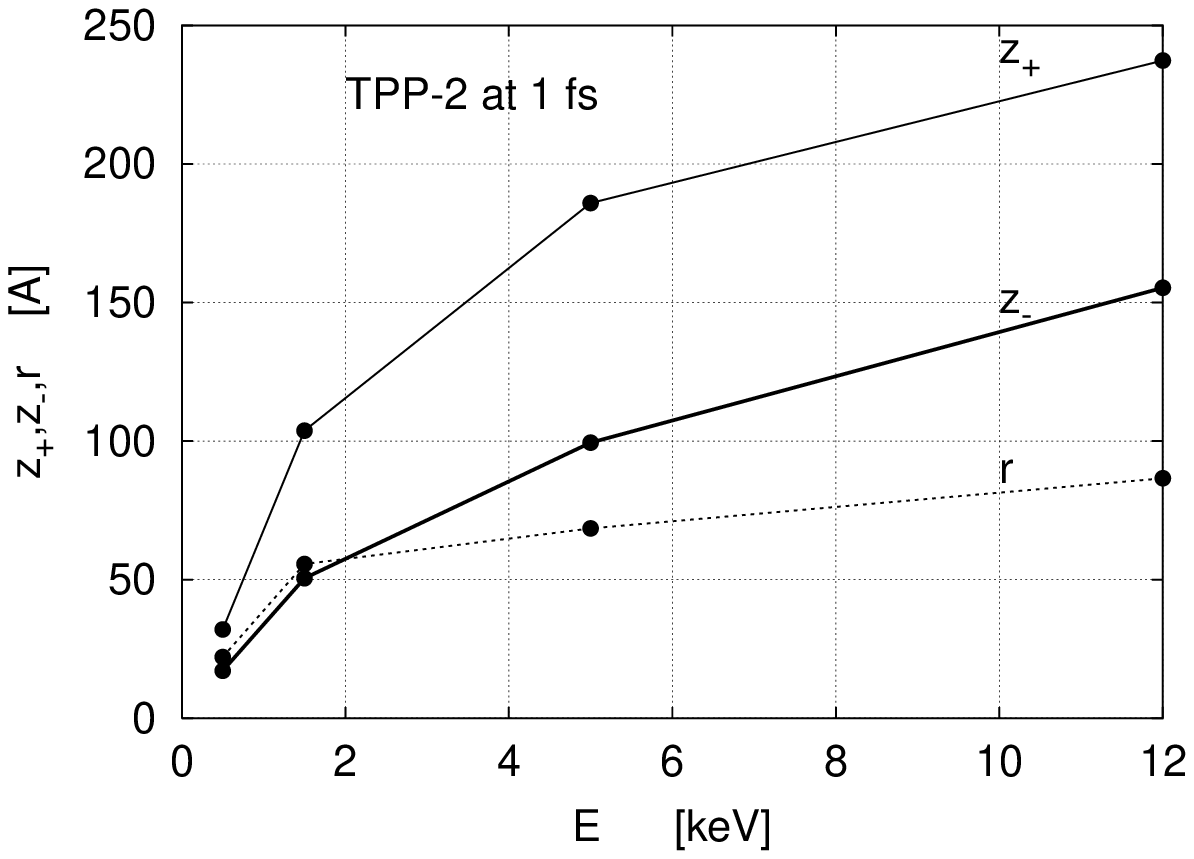}\hfill\mbox{}\\
b)\epsfig{width=8cm, 
file=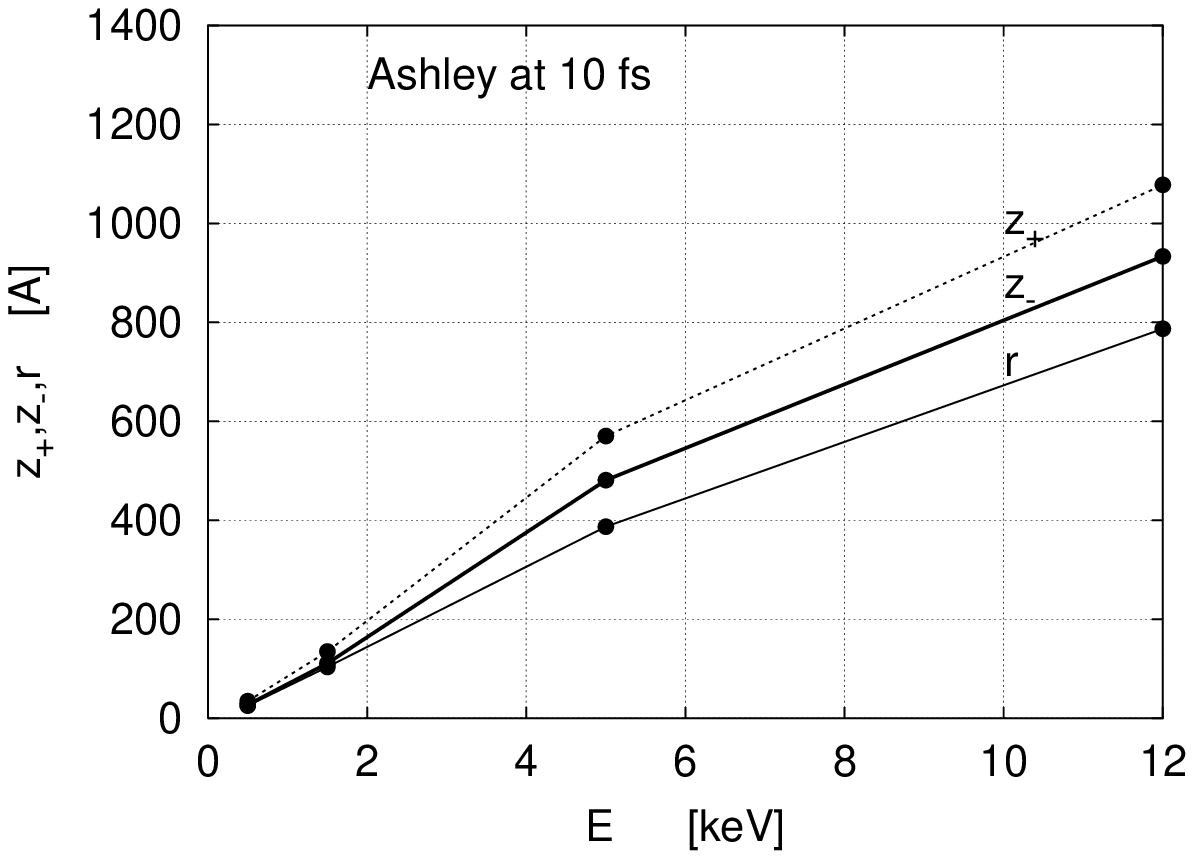}
\epsfig{width=8cm, 
file=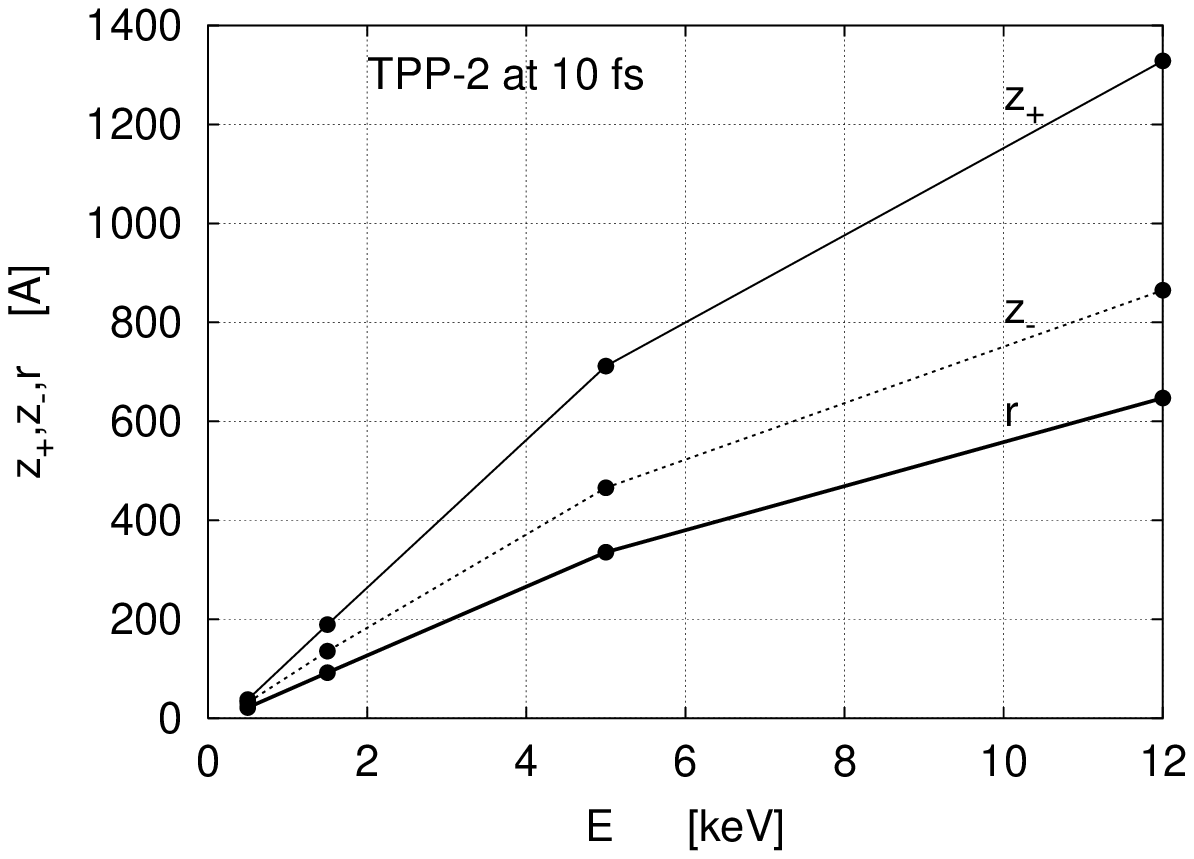}\hfill\mbox{}\\
c)\epsfig{width=8cm, 
file=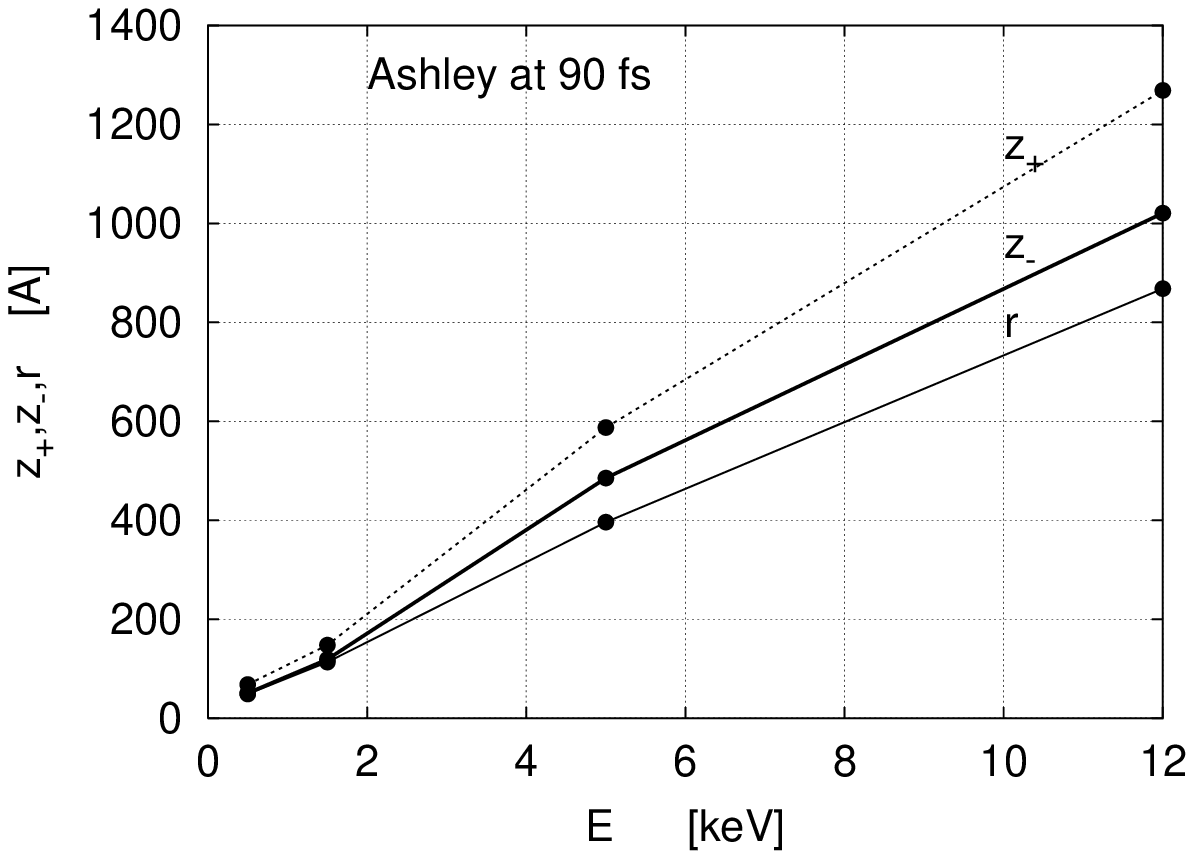}
\epsfig{width=8cm, 
file=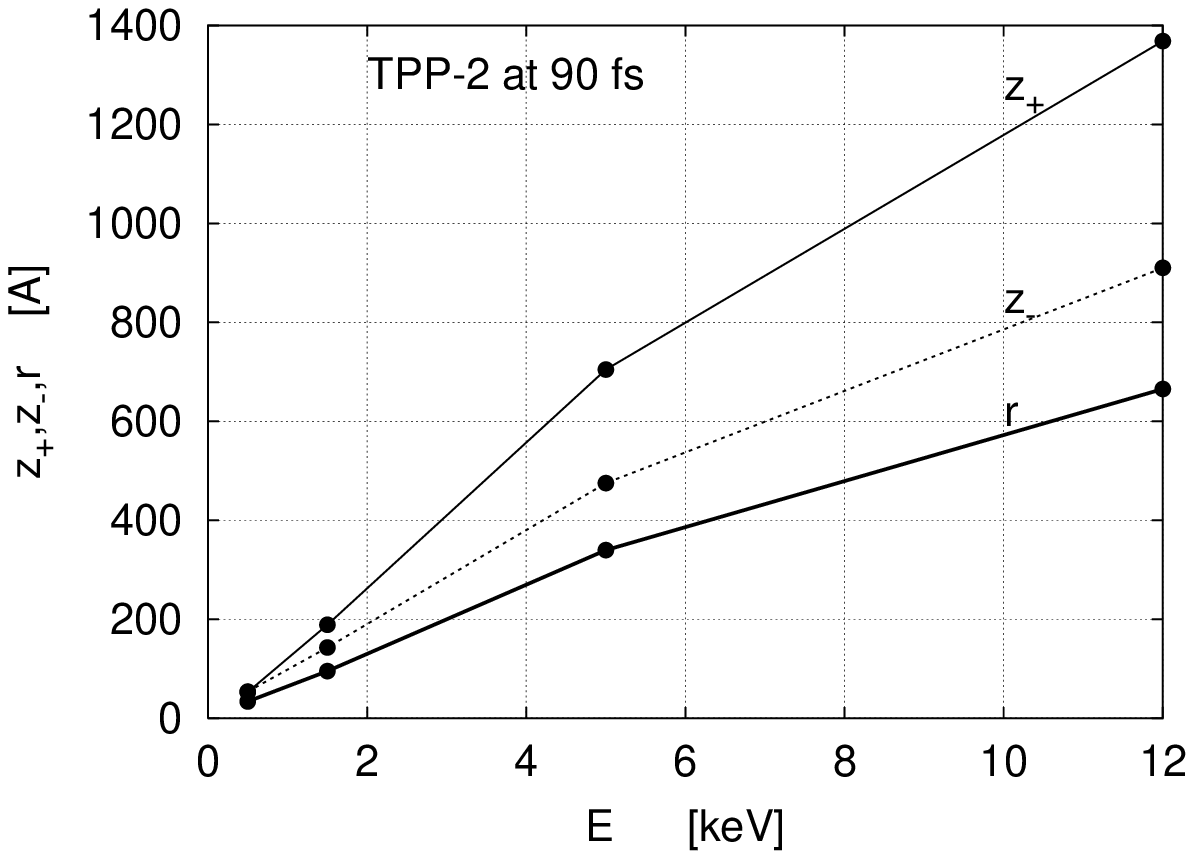}\hfill\mbox{}\\
\caption{{\footnotesize Parameters  $z_{+}$, $z_{-}$ and  $r$ describing 
the spatial structure of the electron cloud.
The data from 500 cascades were collected at times: {\bf (a)} $t=1$ fs, 
{\bf (b)} $t=10$ fs, {\bf (c)} $t=90$ fs at 
primary energies of $E=0.5,1.5,5,12$ keV from Ashley's model and the 
TPP-2 model with {\bf no} energy transfer allowed to the lattice.}
}
\label{f12}
\end{figure}

The final cloud at higher impact energies is pear-shaped. Figure \ref{f12} 
gives a quantitative description of the development of such clouds, using 
parameters shown in Figure \ref{f11}. Parameters $z_+$, $z_-$ and $r$ 
represent root mean square values from 500 simulations,
\eqn
z_+&=&\sqrt{ \langle (z-z_{CM})^2 \rangle },\,\,\, z>z_{CM}\nonumber\\
z_-&=&\sqrt{ \langle (z-z_{CM})^2 \rangle },\,\,\, z<z_{CM}\nonumber\\
r  &=&\sqrt{ \langle (x-x_{CM})^2+(y-y_{CM})^2 \rangle },
\label{zzxy}
\eqnx
and were calculated in respect to the center of mass of the cloud. The results 
confirm that the cloud is elongated along the axis of the primary impact (Z) 
at early times, and becomes more spherical later.


\subsection*{Conclusions} 

Our results can be used to estimate damage by photo electrons 
($E=0.5-12$ keV) 
in diamond and other carbon-based covalent compounds. 
The Monte-Carlo code may be adopted to simulate multiionization
phenomena in different systems, ranging from the
explosion of atomic clusters to the formation of warm dense matter and 
plasmas. The model could
also be used to estimate ionization rates and the spatio-temporal
characteristics of secondary electron cascades in biological
substances. 

%
%
%
%
%
%
%
\noindent
%
%
%
\noindent
%
%
\noindent
%
%
\noindent
%
%
\noindent
%
%
\noindent

\section*{Acknowledgments} 

We are grateful to Gyula Faigel, Zoltan Jurek, Michel A. van Hove 
for discussions. B.\ Z.\ thanks Peter Druck and the Institute of 
Theoretical 
Physics II at the Ruhr University in Bochum for providing the access to 
the
computational unit. This research was supported in part by the Polish 
Committee for Scientific Research with grant No.\ 2 P03B 05119, 
the EU-BIOTECH Programme and the Swedish Research Councils. 
A.\ S.\ is grateful to STINT for a distinguished guest professorship. 
B.\ Z.\ was supported by the Wenner-Gren Foundations.


\end{document}